# In-Memory Database Systems - A Paradigm Shift

Mohit Kumar Gupta[1], Vishal Verma[2], Megha Singh Verma[3]

[1]Blogger at www.vedantatree.com
[2]Department of Computer Science, M L N College, Yamuna Nagar, INDIA
[3]Department of Computer Science, D A V College for Girls, Yamuna Nagar, INDIA

*Abstract* — **In today's world, organizations like Google, Yahoo, Amazon, Facebook etc. are facing drastic increase in data. This leads to the problem of capturing, storing, managing and analyzing terabytes or petabytes of data, stored in multiple formats, from different internal and external sources. Moreover, new applications scenarios like weather forecasting, trading, artificial intelligence etc. need huge data processing in real time. These requirements exceed the processing capacity of traditional on-disk database management systems to manage this data and to give speedy real time results. Therefore, data management needs new solutions for coping with the challenges of data volumes and processing data in real-time. An in-memory database system (IMDS) is a latest breed of database management system which is becoming answer to above challenges with other supporting technologies. IMDS is capable to process massive data distinctly faster. This paper explores IMDS approach and its associated design issues and challenges. It also investigates some famous commercial and open-source IMDS solutions available in the market.**

*Keywords* — **In-Memory Database System (IMDS), Design issues and challenges for IMDS, Commercial and open-source IMDS.**

## I. INTRODUCTION

With the increasing demand of real time data processing, traditional (on-disk) database management systems are in tremendous pressure to improve the performance. With the increasing amount of data, which is expected to touch 40ZB (1ZB = 1 billion terabytes) by 2020, means 5247 GB of data per person [1], and with traditional DBMS architecture, it is becoming more and more challenging to process the data and to produce analytical results in almost real time. For on-disk databases, disk I/O operations are the main bottleneck, which are very slow operations and can't be optimized beyond a limit being mechanical in nature. Although traditional on-disk DBMS have tried to improve on this by introducing various caching techniques to cache the frequently accessed data, however, it comes at the cost of synchronization of cache with disk and vice versa and to implement various complex logic to manage transaction and resources, which itself pose as a limitation to performance. So what is the way forward?

Here comes the in-memory database system concept, which actually changed the whole architecture paradigm for the database management system. An in-memory database system or main-memory database system is a breed of database management system that stores data entirely in main memory instead of keeping it on disk [2]. With decreasing cost of main memory, and advance technological innovations, it becomes quite feasible to store large amount of data in main memory.

Once data is stored in main memory, speed of reading and writing the data will be improved drastically as it eliminates disk I/O operations. A POC done by McObject shows; that in-memory database supports read at 4x speed and write at 420x speed than traditional DBMS [3]. It is a big d3fference if we can transform this benefit in real time data processing and hence in managing and processing the big data. Further, having all data in main memory, now there is no need to implement complex caching logics and hence caching overhead is also eliminated. This is how IMDSs are moving to win over traditional databases for speed challenges.

Applications of in-memory databases will be in all domains that require real-time performance and very low latency like weather forecasting, trading application, social networking websites, artificial intelligence etc. Another important use for in-memory database systems is in real-time embedded systems such as IP network routing, telecom switching, industrial control etc.

## II. ARCHITECTURE OVERVIEW

Main architectural attribute in IMDS is that whole design is geared towards using the main memory for data storage, instead of using disks. Fig. 1 shows a very simply high level design of IMDS.

This is a major paradigm shift in DBMS design approach, which triggers many other design optimizations. Now DBMS needs not to be worried about optimizing the disk I/O operations, and about caching like techniques for these optimizations. Whole designed will be geared to have high performance data access, manipulation and analysis relying on main memory data store. Since IMDS involves no disk oriented algorithms, therefore it frees the database design to leave typical B-Tree kind of data structure and go for better main memory access friendly data structures like T-Tree, and possibly open the arena to invent better in-memory storage data structures. Further, Query optimization is now focused on improving the in-memory data structure and algorithms to execute the query, instead of improving the I/O kind of operations, which are having mechanical limitations. Overhead to manage the concurrent transactions will be lesser, because data access is much faster and hence the locks will be freed comparatively faster.

As whole data is loaded in-memory, so the distributed data management nodes can also use the shared main memory locations or high speed WAN network to enable the virtually one data location and hence faster data access even in case of distributed nodes [4]. Moreover, having all data in main

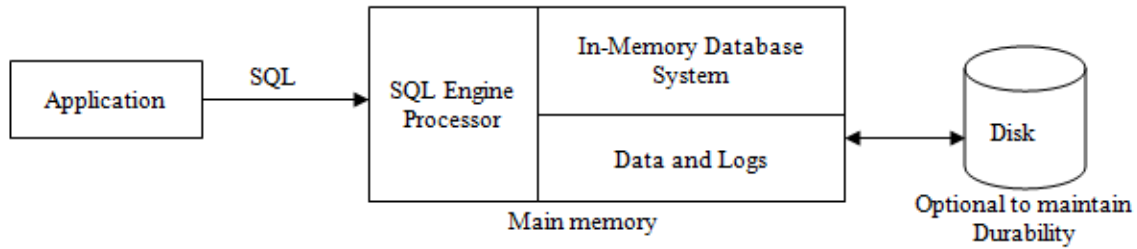

Fig. 1 High Level Design of IMDS

memory enable the design to utilize the extended capacity of memory pointers on 64 bit systems.

A common misconception is that traditional database systems can give the same performance if we change the data store from disk to main memory. And that IMDS is nothing but a traditional database having data store in RAM only. However, it is not true. Reason is that it is not only the data access operations which are getting fast with main memory based design, but there are many other optimizations which are possible now. As discussed earlier, now database systems need not to worry about complex caching and synchronization algorithms. Further, we can redesign and optimize data structures, indexes approach, query execution rules and algorithms considering main memory data store; which would be quite different than traditional databases [3].

*A. Few Design Highlights*

1) *Client/ server architecture:* Some IMDSs use client/ server architecture, where server can be run on powerful machines with multi-core processors and vast amounts of RAM and multiple clients can interact with server for requests. Even when residing on the same computer, client and server processes communicate via inter-process communication messaging. Other IMDSs use an in-process architecture in which the database system runs entirely within the application process; instead of separate client and server processes. Later is useful in especially useful in embedded applications. In-process architecture is simpler, which means it has a smaller code size (shorter execution path). Simpler code is also less prone to defects. In-process design eliminates inter-process communication, resulting in lower latency. In-process IMDSs further accelerate performance and reduce complexity by eliminating server tasks such as managing sessions and connections, and allocating and de-allocating resources. However former, i.e. Client/server is an enterprise DBMS kind of design. It is useful in right sizing the database capacity and performance by installing the server software on a more powerful computer, and clients may be on lesser powerful nodes. As a result, it is capable to manage resources in better way and serve requests in optimized manner even with limited resources.

2) *Shared Nothing architecture (SN):* The 'Shared nothing architecture' principle is that every database node work absolutely independently without having any dependency on other nodes. In case of dependency, failure at one node can affect other nodes also and hence bring down all or most of the database system. SN helps in achieving high availability in case of node failure, as control can be routed to other active node without any dependency.

3) *Partition aware databases:* It is an approach where data storage is managed by horizontal partitioning i.e. tables are partitioned in rows and placed in different servers where these are mostly required. One of criteria for partition could be based on demography. Use of horizontal partitioning results in reduced number of rows in a table and hence less overhead of indexing and searching. It also means having data in a node which is near to its consumer. However, there could be a problem when some operation needs joins on partitioned table and other tables. Now if other tables or data are present in different node, a distributed join will be slower. To mitigate this problem, a design approach is to make database aware that which tables or data should be collocated on a node. Once database is aware about it, it will make sure that related data should be stored on single node (replication may include replicating whole data set to other nodes to avoid single point of failure). Hence, now the joins will take place on single node and much faster without the overhead of distributed data.

4) *Scalable infrastructure:* Having SN architecture in place, it is possible for any node from available set to cater the request. Hence, it is also easily possible to add any number of nodes as and when required. Architecture can support the automatic replication of data to this newly added node and hence making it fully functional to cater the requests. In the same way, nodes can also be removed easily if load is decreased. The high availability distributed architecture will automatically ensure that no new request is being sent to this node and will be redirected to other available nodes only. All this means, a huge saving in infrastructure cost if hosting platform is cloud, where provisioning of resources to scale up or down can be very fast on demand. This also means that DB can scale linear just by adding more nodes horizontally, which is much easy than adding or removing capacity vertically as in case of traditional database systems.

5) *Disaster recovery:* Being highly available systems, it is very easy to make disaster recovery itself. When data is available on multiple nodes, it becomes obvious to recover any failed node from other available nodes. However, another design principle applied is to provide disk storage backup for systems logs. Here backup is done in asynchronous processes so that it won't affect the main database performance. Users may be able to choose the backup options from various available options like backup to disk on selective nodes, replication to other nodes and may be more in future.

6) *Distributed data with WAN based clustering support:* For a worldwide application, different part of database would be required in different geography. One of the design possibilities in IMDS is to partition the data based on data required at a particular node. So node will host the data which is most required there. Other commonly accessed data tables will be replicated in real time basis. However, in case, if data hosted at other nodes is also required for some specific scenarios, all the nodes can act as single database by exploiting the high speed network power of LAN and customer will not feel the difference of having distributed database across the nodes. This is another valuable design element, which can help in managing the storage capacity and to manage the amount of data at a particular node.

*B. Challenges for IMDS*

Despite the promising design attributes for high performance, IMDS face some challenges in compare to traditional on-disk databases. While the robust traditional database systems, as a matter of design based on hard disk, guarantee atomic, consistent, isolated and durable (ACID) nature of the transactions. IMDSs have various challenges with assuring the durable characteristic due to its RAM-based design. Different products are offering diverse ways to cope with this issue. To achieve durability, in-memory database systems are applying the following techniques [5]:

1) *Persistent main memory*: Latest technological developments in main memory arena like NVRAM (Non-Volatile RAM) or battery powered main memory enable the scenario where data in main memory will not be lost.

2) *Transaction logging:* With transaction logging, every transaction will be logged to a persistent store, which will enable the roll-forward mechanism to restore the database in case of power failure. To avoid making transaction logging the bottleneck in overall performance, different design strategies are to write the logs first in stable memory (faster than disk write) and then to disk in asynchronous processes and hence the main database operations will not wait for the logging to be completed.

3) High *availability implementation:* High availability implementation is another design approach to address this problem. High availability means replicating the data to other available nodes in real time. That will enable to create the copies of data in almost real time and hence will reduce the probability of complete data failure. In case, if any one node goes down, other node will take over the request and will process it.

So various approaches given above are answer to the challenge of volatile main memory and this list is increasing very fast with more and more design and technical innovations in this area.

III. IMDS SOLUTIONS AVAILABLE IN MARKET

In recent years, variety of IMDS solutions (both commercial and open source) made available in market from giant players of database market such as IBM, Oracle, SAP, VMWare etc. Although all of them share the capability to maintain the database in main memory and supports industry standards such as SQL for data processing, they offer different set of features. Following sub-sections investigates some famous commercial and open-source in-memory databases:

*A. Commercial IMDS*

1) *TimesTen*: TimesTen [6] is in-memory relational database system from Oracle with features like durability, query optimization, recoverability etc. It offers instant responsiveness and very high throughput required by today's real time applications such as telecom, capital markets and defence. It also provides multiple interfaces such as JDBC, ODBC and other SQL APIs.

2) *SolidDB:* SolidDB [7] is a hybrid disk/ in-memory relational database system from IBM. It offers extreme speed, availability and adaptability required for mission-critical applications. There are a number of deployments of solidDB in telecommunication networks, enterprise applications and embedded software and systems.

3) *extremeDB:* extremeDB [8] from McObject is an extremely fast in-memory database system. It is designed explicitly for real-time applications and for embedded systems such as set-up boxes, telecom/ networking devices, industrial control system etc. It states to offers unmatched performance, reliability and development flexibility.

4) *SQLFire:* SQLFire [9] is an in-memory distributed SQL database from VMware vFabric. It states to offer high throughput, dynamic & linear scalability, and continuous availability of data. It is designed by utilizing the research and development done on in-memory cache 'GemFire' by same vendor and hence is quoted to be very mature and feature rich.

5) *HANA:* HANA [10] is distributed in-memory relational database system from SAP. It supports features like column-based storage and queries, data compression and parallel processing which makes possible forecasting, planning, analysis and simulation in real or near to real time. It states to offer support for complex queries and high performance for complex queries.

*B. Open-source IMDS*

1) *SQLite*: SQLite [11] is an open-source relational in-memory database engine. It is small, fast and reliable DBMS suited for embedded systems such as cellphones, PDAs, set-up boxes etc. It is also used for local/ client storage in web browsers. However, it has limited support for complex SQL queries, triggers and views.

2) *CSQL:* CSQL [12] is another open-source main-memory relational database system developed at sourceforge.net. It is designed to provide high performance on simple SQL queries and DML statements that involve only one table. It supports to work in embedded as well as client/server mode. Apart from acting as relational storage engine, it can also be used as a cache for existing disk-based commercial databases.

*3) MonetDB:* MonetDB [13] is an open-source column-oriented main-memory database management system developed at the National Research Institute for Mathematics and Computer Science in the Netherlands. It was designed to provide high performance on complex queries in large databases, such as combining tables with hundreds of columns and multi-million rows. MonetDB is one of the first database systems to focus its query optimization. It has been successfully applied in high-performance applications for data mining, OLAP, XML Query, GIS etc.

## IV. CONCLUSIONS

When it comes to processing large amounts of data with low latency, in-memory database seems to be a definitive answer and a turning point in design strategies. They offer so many advantages and it is only a matter of time when we shall see many new innovations to address their current challenges or limitations. With in-memory replicated data nodes, IMDS provides a very efficient way to support highly available and performance oriented database management systems. With continuous improvements in main memory technology and high speed networks, future will support the 'store anywhere, use anywhere' with high processing speed and virtually one database base system accessible anywhere in world. If all knowledge accumulated in world can be processed and used at high speed, it will be a definitive revolution in many high-tech fields like Artificial Intelligence, various forecasting services and so on.

## AUTHORS

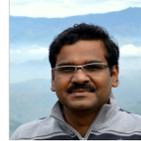

Mohit Kumar Gupta is a Software Professional having 12+ years of experience with various Organizations in Financial Services, Public Finance and IT Services domains. He is having Masters degree in Computers Applications and another Masters degree in Business Administration. His interest areas are Open Sources, Framework Development, Research in new technologies, Architecture and Designing. He is blogger at http://www.vedantatree.com. His complete profile can be found at http://www.linkedin.com/in/mohitkgupta.

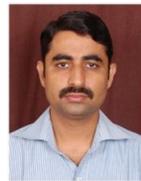

Vishal Verma is an Assistant Professor at Department of Computer Science, M L N College, Yamuna Nagar, Haryana (INDIA). He received MCA in 2001 from Kurukshetra University, Kurukshetra (INDIA) and M Phil (Computer Science) in 2008 from Madurai Kamaraj University, Madurai (INDIA). His total teaching experience is more than 12 years and is presently pursuing PhD (Computer Science) at Maharishi Markandeshwar University, Mullana, Ambala (INDIA). His current research focus is on Rendering Techniques, Image Processing and latest trends in Database Systems.

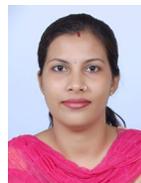

Megha Singh Verma, received M.Tech. degree in Computer Science and Engineering from Maharishi Markandeshwar University, Mullana, Ambala (INDIA) in 2011. She is presently working as Assistant Professor at Department of Computer Science, D A V College for Girls, Yamuna Nagar, Haryana (INDIA). She has more than 2 years of teaching experience and her areas of interest are Database Systems and Computer Graphics.